\newcommand\tf{\scriptsize}

\documentstyle{epsf}

\documentstyle{l-aa}

\begin{document}

\thesaurus{13 
  (03.13.2; 
  03.20.8; 
  08.06.3)} 

\title{Automated classification of IUE low-dispersion spectra}

\subtitle{I.  Normal Stars}

\author{E.\ts F. Vieira\inst{1} \and J.\ts D. Ponz\inst{2}}

\institute{LAEFF
           \and
           ECNOD/ESA \\
           Villafranca, P. O. Box 50727, 28080 Madrid, Spain\\
           efv@laeff.esa.es, jdp@vilspa.esa.es}

\offprints{ J.\ts D.  Ponz }

\date{ Received November 30th, 1994; accepted January 4th, 1995 }

\maketitle

\begin{abstract}

  Along the life of the IUE project, a large archive with spectral
  data has been generated, requiring automated classification methods
  to be analyzed in an objective form.  Previous automated
  classification methods used with IUE spectra were based on
  multivariate statistics.  In this paper, we compare two
  classification methods that can be directly applied to spectra
  in the archive: metric distance and artificial neural
  networks.  These methods are used to classify IUE low-dispersion
  spectra of normal stars with spectral types ranging from O3 to G5.
  The classification based on artificial neural networks performs
  better than the metric distance, allowing the determination of the
  spectral classes with an accuracy of 1.1 spectral subclasses.

   \end{abstract}

\section{Introduction}

The availability of large spectral archives and the efficiency
achieved with modern instrumentation require automated classification
methods to improve the classification by visual inspection.  These
methods are still in exploratory phase; the aims are clear, to devise
an objective, repeatable and robust classification scheme providing an
estimation of systematic and random errors, and allowing the
quantification of spectral resolution and signal to noise ratio on
classification errors.

Automated classifiers of stellar spectra can be divided into metric
distance algorithms, multivariate statistics and artificial neural
networks (hereafter ANN).  Metric distance methods were originally
proposed by Kurtz and LaSala (Kurtz \cite{kurtz82}, \cite{kurtz84},
LaSala \cite{lasa}) and have been used by Penprase (\cite{pen}) to
classify stellar spectra using the digital spectral atlas of Jacoby et
al. (\cite{jac:etal}) as template.  Multivariate statistical methods
are linear algorithms used for exploratory data analysis.  These
methods have been applied to spectral classification by using
Principal Component Analysis (PCA) to reduce the dimension of the
problem, followed by Cluster Analysis (CA) to discover groups of
objects in the parameter space obtained in the previous step (Murtagh
\& Heck \cite{mur:heck} and references herein).  Stellar
classification with ANN is a new approach that has been used by von
Hippel et al. (\cite{hip:94a}, \cite{hip:94b}) to confirm the visual
classification of the Michigan Spectral Catalogue on objective prism
spectra, determining the temperature classification to better than 1.7
spectral subclasses from B3 to M4.  Gulati et al. (\cite{gula:etal})
classify the spectral atlas of Jacoby et al. (\cite{jac:etal}) with an
accuracy of 2 spectral subclasses, based on selected spectral
features.

The availability of the IUE low-dispersion archive (Wamsteker et al.
\cite{wam:etal}) allows the application of pattern recognition methods
to explore the ultraviolet domain.  The analysis of this archive is
especially interesting, due to the homogeneity of the sample.  As
indicated by Heck (\cite{heck:87}), it is important to remember at
this point that MK spectral classifications defined from the visible
range cannot simply be extrapolated to the ultraviolet spectral range.
So far, only multivariate statistical methods have been used with IUE
spectra.  Egret and Heck (\cite{egret:heck}) analyze the relative
fluxes at 16 selected wavelengths of O and B stars with PCA.  Egret et
al. (\cite{egret:etal}) analyze low-resolution spectra using PCA on 93
variables computed as median flux values at certain wavelength bands
and selected absorption and emission lines.  These analyses indicate a
high correlation between the first principal component and the
temperature.  Heck et al. (\cite{heck:etal}) classify the IUE
Low-Dispersion Spectra Reference Atlas (Heck et al. \cite{heck:84}) of
normal stars.  Weighted intensities of sixty lines together with an
asymmetry coefficient describing the continuum shape are used for the
classification.  The algorithm consisted of PCA followed by CA to
define different groups that confirmed the manual classification of
the Atlas.  Imadache and Cr\'ez\'e (\cite{imadache:creze}) and
Imadache (\cite{imadache}) extend the sample and generalize the
method, using the full spectral range instead of pre-selected spectral
features.

The present work has been done within the context of the IUE Final
Archive project.  The aim is to provide an efficient and objective
classification procedure to explore the complete IUE database, based
on methods that do not require prior knowledge about the object to be
classified.  Two methods are compared: a simple metric distance
method, and a supervised ANN classifier.  The input sample and data
preparation steps are described in Sect.~2.  The classification method
based on metric distance is summarized in Sect.~3.  Section 4 explains
the classification method using ANN.  The results obtained with both
methods are described in Sect.~5, and the conclusions are presented in
Sect.~6.

\section{The sample spectra}

The spectra were taken from the IUE Low-Dispersion Reference Atlas of
Normal Stars (Heck et al. \cite{heck:84}, hereafter the Atlas),
covering the wavelength range from 1150 to 3200 \AA.  The Atlas
contains 229 normal stars distributed from the spectral type O3 to K0.
The classification given in the Atlas was carried out following a
classical morphological approach (Jaschek \& Jaschek
\cite{jaschek:2}), based on UV criteria alone.  The set of 64 standard
stars selected in the Atlas, with spectral types from O3 to G5, was
used as a template in the metric distance classification and was the
training sample in ANN classification.  The test set contained 163
spectra, excluding the 64 standard stars and two stars with spectral
types G8 and K0, outside the spectral types covered by the training
set.

The spectra were obtained by merging together data from the two IUE
cameras, sampled at a uniform wavelength step of 2 \AA, after
processing with the standard calibration pipeline.  Although the
spectra are good in quality, there are two aspects that seriously
hinder the automated classification: interstellar extinction and
contamination with geo-coronal Ly-$\alpha$ emission.  Some
pre-processing was required to eliminate these effects and to
normalize the data.

All spectra were corrected for interstellar extinction by using
Seaton's (\cite{seaton}) extinction law.  Due to the properties of the
extinction law at $\lambda_1 = 1600$, $\lambda_2 = 2400$ and
$\lambda_c = 2175$ \AA\, the color excess $E(B-V)$ was estimated as

\begin{equation}
  \label{eq:ebv} E(B-V) = 1.368 \log \left[ \frac{225 f^o_1 + 575
f^o_2}{800 f^o_c} \right] \,.  \end{equation}

\begin{figure}[htbp]
  \epsfxsize=8cm \epsfbox{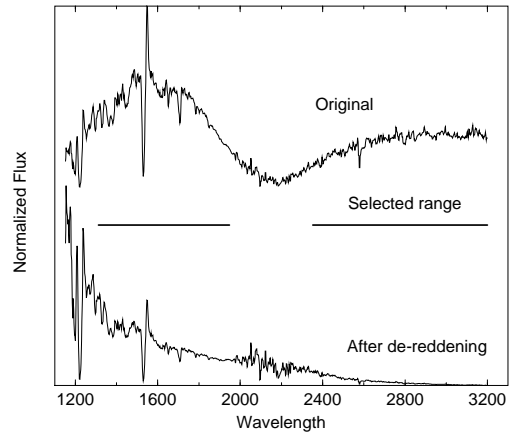}
  \caption{Original (top) and de-reddened (bottom) spectra
    corresponding to a O4 star.
    The selected range is indicated by the solid lines in the middle}
  \label{fig:dered-2} \end{figure}

The observed fluxes $f^o_1$, $f^o_2$ and $f^o_c$ were obtained by
filtering the high frequency components in the transformed Fourier
space (LaSala and Kurtz \cite{lasa:kurtz}).  Figure \ref{fig:dered-2}
shows a typical O4 spectrum before and after correction for reddening.

The region below 1250 \AA\ was excluded from the analysis, to
eliminate the geo-coronal Ly-$\alpha$ component, and also the spectral
band from 1950 to 2350 \AA, because of the low signal to noise ratio
in this region.  The selected wavelength range is indicated by the
solid lines in Fig.~\ref{fig:dered-2}.  The resulting spectra
contained $N = 744$ flux values, which were normalized to obtain a
mean flux value of zero and a sum of the absolute values of the
normalized fluxes equal to one.

\section{Classification using metric distance}

Normalized spectra are considered vectors in $\bbbr^N$ and a metric is
introduced in the vector space.  In this classification scheme the
metric distance between the object spectrum and each spectrum in the
training set is computed and the spectral class of the star in the
training set having the minimum distance is assigned to the object.

Let $f_{ij} = f_i(\lambda_j);\; j = 1, \ldots, N$ be the flux of the
$i$-th star in the catalogue and $s_{kj} = s_k(\lambda_j);\; j = 1,
\ldots, N$ be the flux of the $k$-th standard star in the training
set, after correction for reddening and normalization.  The distance,
$d_{ik}$, is defined by \begin{equation}
  \label{eq:md}
  d_{ik}^2 = \frac{1}{N} \sum_{j=1}^N (f_{ij} - s_{kj})^2 \,.
\end{equation}

\section{Classification using ANN}

A supervised classification scheme based on artificial neural networks
(ANN) has been used.  This technique was originally developed by
McCullogh and Pitts (\cite{mc:pitts}) and has been generalized with an
algorithm for training networks having multiple layers, known as
back-propagation (Rumelhart et al. \cite{rum:etal}).  In a general
form, a neural network represents a function, $F$, that maps a given
input set into a selected output set.  Assuming that normalized
spectra are elements in a $N$-dimensional vector space and spectral
classes are defined by $M$-dimensional classification vectors, the
network approximates the mapping $F: \bbbr^N \rightarrow \bbbr^M$, in
such a way that a standard star, $\vec{s\/}_k$, is associated with the
vector $\vec{c\/}_k = F(\vec{s\/}_k)$.  The output vector defines the
spectral class so that O0 is given by $(1, 0, 0, \ldots, 0)$, O1 is
represented by $(0, 1, 0, \ldots, 0)$ and so on.  In this form, the
network can be regarded as a classifier that maps the input space of
normalized fluxes into $1\ of\ M$, i.e., one output unity and all
others zero.  Such a network, with a squared-error cost function,
gives a good estimation of Bayesian probabilities (Richard \& Lippmann
\cite{rich:lipp}), so that for an input spectrum $\vec{f}$, the $i$-th
component of the output vector is the probability $P(C_i|\vec{f})$ for
class $i$ given the input spectrum.

The supervised classifier works in two phases: During a first step,
the learning phase, the classifier is trained with the standard stars
in the Atlas together with the associated output vectors defining the
spectral classes.  In a second step, the test sample is presented to
the network and spectral types are assigned as defined by the maximum
value of the probability distributions estimated by the classifier.

The network architecture consists of an input layer, one or more
hidden layers and the output layer.  The input layer contains $N$
nodes that accept the individual components of the input vector and
distribute them to the nodes in the second layer.  Nodes in a layer
receive the weighted sum of the output from all the nodes in the
previous layer, so that the input of node $j$ is \begin{equation}
  \label{eq:rum-1}
  x_j = \sum_i y_i w_{ji}\,, \end{equation} where $y_i$ is the output
of node $i$ in the previous layer and $w_{ji}$ is the weight
associated to the connection of node $i$ to node $j$.  The output of
node $j$ is computed using the sigmoid transfer function
\begin{equation}
  \label{eq:rum-2}
  y_j = {1\over 1 + {\rm e}^{-x_j}}\,.  \end{equation}

During the training phase, input vectors with normalized fluxes of the
standard stars are presented to the network and Eqs.~(\ref{eq:rum-1})
and (\ref{eq:rum-2}) are applied in a feed-forward mode, until the
output layer with $M$ nodes is reached.  For each star in the training
set, the output vector $\vec{o}$ is compared with the desired
classification vector $\vec{c}$, defining the spectral type of the
standard star, and the error is evaluated as \begin{equation}
  \label{eq:rum-3}
  E = {1\over 2} \sum_{j=1}^M (o_{j} - c_{j})^2\,.  \end{equation}

The minimization of the error is achieved during the training phase by
changing the connection weights according with the error feedback
mechanism, known as back-propagation (Rumelhart et
al. \cite{rum:etal}).  During this step, connection weights $w_{ji}$
are updated using the rule: \begin{equation}
  \label{eq:rum-9}
  \Delta w_{ji}(t) =
   -\eta{\partial E\over\partial w_{ji}(t)}+\alpha\Delta
w_{ji}(t-1)\,, \end{equation} where $\eta$ is the learning rate,
$\alpha$ is the momentum factor to reduce oscillations during the
learning process, $t$ is the iteration number and $E$ is the error,
given by Eq.~(\ref{eq:rum-3}).

The procedure was repeated for all spectra in the training set in
several iterations.  The training set was sampled randomly before each
iteration, to avoid trends during the learning phase.  After each
iteration, the average error for all the stars in the training set was
computed using Eq.~(\ref{eq:rum-3}), to control the convergency of the
procedure.  After 2000 iterations there was no substantial improvement
in the classification.  To prevent excessive corrections during the
first iterations, the parameters $\eta$ and $\alpha$ were linearly
increased with the iteration number, reaching the operational values
of 0.1 and 0.5, respectively, after 100 iterations.

\begin{figure}[htbp]
  \epsfxsize=8cm
  \epsfbox{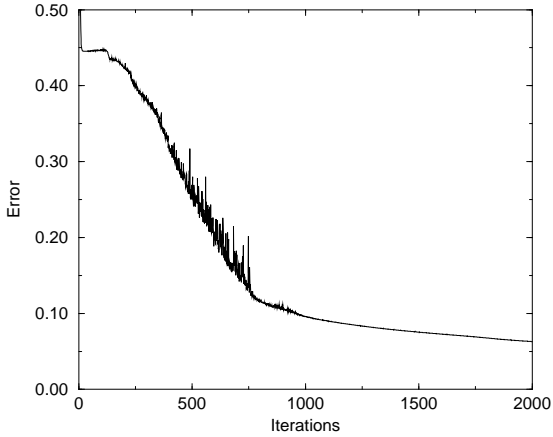}
  \caption{Average error as a function of the iteration number for ANN
           classifier with $744 \times 120 \times 120 \times 51$
topology}
  \label{fig:e-total} \end{figure}

In our study we have used several network configurations, with an
input layer consisting of $N = 744$ nodes, corresponding to the number
of spectral flux values, one or two hidden layers and an output layer
having $M=51$ output nodes.  A total of six different network
topologies were used in the study with 30, 60 and 120 nodes in the
hidden layers.  In addition, different output distributions were used
during the training phase, by convolving the discrete $\delta$
functions representing the class probabilities with gaussian
distributions having different standard deviations. The best results
were obtained for a standard deviation of $0.7$.  Table~1 summarizes
the classification statistics for each configuration, the first two
columns define the network topology, the third column shows the
correlation coefficient, $r$, between the classification obtained with
ANN and the manual classification given in the Atlas, and the fourth
column is the standard deviation, $\sigma$, of the differences between
the two classifications.  The network topology $744 \times 120 \times
120 \times 51$ produces the best classification.  This network was
used in the analysis below.  Figure~\ref{fig:e-total} shows the total
error, given by Eq.~(\ref{eq:rum-3}), as a function of the number of
iterations for the $744 \times 120 \times 120 \times 51$ topology.

\begin{table} \label{tab:conf} \caption[]{Network Configurations}
\begin{flushleft}
  \begin{tabular}{cccc}
    \hline\noalign{\smallskip}
    Hidden & Hidden && \\
    Nodes & Layers & $r$ & $\sigma$ \\
    \noalign{\smallskip}
    \hline\noalign{\smallskip}
       120 & 2 & 0.988 & 1.107 \\
        60 & 2 & 0.983 & 1.350 \\
        30 & 2 & 0.946 & 2.379 \\
       120 & 1 & 0.984 & 1.313 \\
        60 & 1 & 0.986 & 1.221 \\
        30 & 1 & 0.982 & 1.378 \\
    \noalign{\smallskip}
    \hline
  \end{tabular} \end{flushleft} \end{table}

Both methods were applied to the test set of 163 stars in the Atlas,
excluding the 64 standard stars.  In the metric distance method, the
class of a test star was determined by the minimum distance given by
Eq.~(\ref{eq:md}), there is no way to assess the quality of the
classification for a given star.  In the classification using ANN,
fluxes of a test star are mapped into the classification space,
producing an output vector that estimates the Bayesian probabilities
for each class.  The spectral type is defined by the maximum value of
this vector.  In addition, the probability distribution provides
information on the quality of the classification for each test star.

\begin{figure}[htbp]
  \epsfxsize=8cm
  \epsfbox{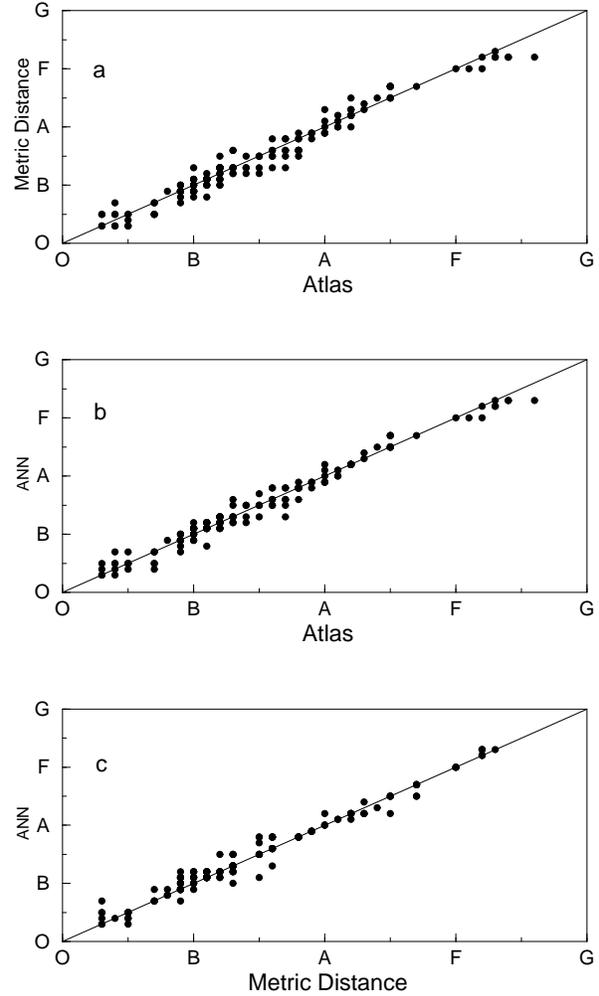}
  \caption{Results of classification: Metric Distance versus Atlas
(top),
    ANN versus Atlas (middle) and ANN versus Metric
    Distance (bottom)}
  \label{fig:results1} \end{figure}

Figures~\ref{fig:results1}a and \ref{fig:results1}b display the
correlation between the manual classification given in the Atlas, in
horizontal axes, and the classifications obtained with the metric
distance and ANN, in vertical axes, respectively.
Fig.~\ref{fig:results1}c shows the correlation between metric distance
and ANN to demonstrate the consistency of both classification methods.
A line of slope unity is also plotted.  These figures show a good
agreement between automated methods and manual classification,
confirming the classification in the Atlas.  However, the results
obtained with ANN are better than the classification with metric
distance.

\begin{table} \label{tab:perf} \caption[]{Comparative Performance}
\begin{flushleft}
  \begin{tabular}{cccc}
    \hline\noalign{\smallskip}
    & \tf{Metric Distance} & \tf{ANN} & \tf{ANN vs.} \\
    Param. &\tf{vs. Catalog}&\tf{vs. Catalog}&\tf{Metric Distance}\\
    \noalign{\smallskip}
    \hline\noalign{\smallskip}
    $\sigma$ & 1.375 & 1.107 & 1.144 \\
    $r$ & 0.982 & 0.988 & 0.988 \\
    \noalign{\smallskip}
    \hline
  \end{tabular} \end{flushleft} \end{table}

\begin{figure}[htbp]
  \epsfxsize=8cm \epsfbox{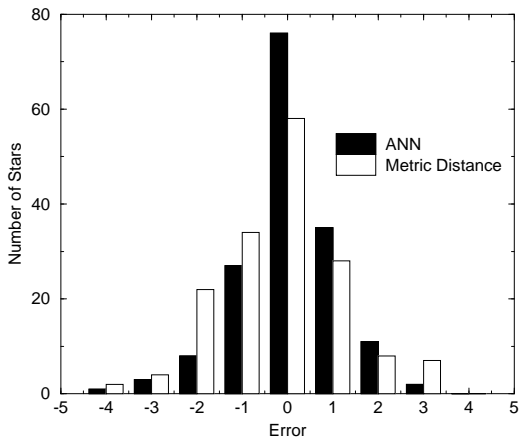}
  \caption{Distribution of classification discrepancies for
    ANN and Metric Distance versus Atlas}
  \label{fig:histo} \end{figure}

Correlation analysis was used to evaluate the performance of the two
classification methods.  Table~2 summarizes the results, including the
standard deviation, $\sigma$, and the correlation coefficient, $r$.
The table shows the superior accuracy of the classification obtained
with ANN, $\sigma = 1.1$, over the metric distance, $\sigma = 1.4$.
Further analysis of the distribution of the classification errors
indicates that 46.6~\% of the stars were correctly classified by ANN,
i.e., in agreement with the spectral class in the Atlas, while only
35.6~\% were correctly assigned using metric distance.  Only six stars
were classified by ANN with a discrepancy larger than 2 spectral
subclasses.  Detailed analysis of these objects show a remarkable
agreement between both methods. In addition, the classification
vectors generated by ANN for these stars have well defined maxima.
Figure~\ref{fig:histo} shows the histogram of the deviations between
the classifications produced by ANN and metric distance and the
classification given in the Atlas. The superior performance of ANN is
evident in these distributions.

\begin{figure}[htbp]
  \epsfxsize=9.5cm
  \epsfbox{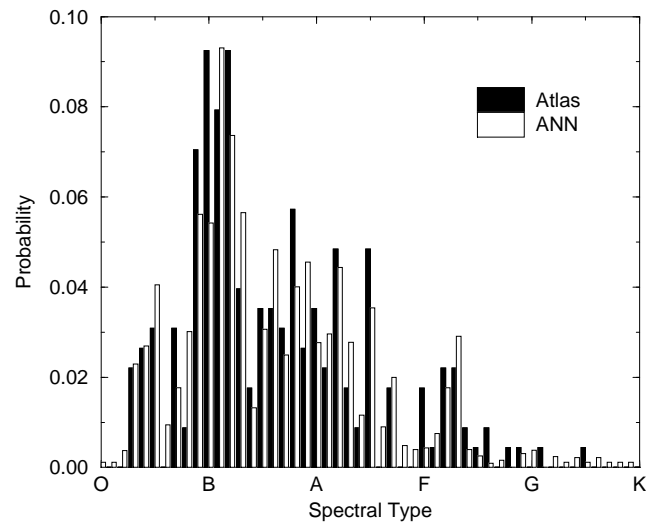}
  \caption{Probability distributions of the spectral classes in the
    Atlas and estimated by ANN}
  \label{fig:dist} \end{figure}

To verify that the ANN classifier gives a good estimation of Bayesian
{\em a posteriori} probabilities, we have analyzed the output
distributions obtained for the complete Atlas.  As a first test, the
sum of the output distributions should be 1 for each sample star.  The
estimated mean value for the total set ($K=227$) is \begin{equation}
  \label{eq:prob1}
  {1 \over K} \sum_{k=1}^K \sum_{i=1}^M o_{ik} = 0.96 \pm 0.06 \,.
\end{equation} As a second test, the output distribution averaged over
the test sample should be equal to the {\em a priori} probability
distribution in the Atlas, $P(C_i)$.  Figure~\ref{fig:dist} compares
the probability distribution in the Atlas with the averaged output
distribution assigned by ANN, estimated as \begin{equation}
  \label{eq:prob2}
  {1 \over K} \sum_{k=1}^K P(C_i | \vec{f\/}_k) =
  {1 \over K} \sum_{k=1}^K o_{ik} \,.  \end{equation} There is a
remarkable agreement between both distributions.  The probability
values obtained from the Atlas and estimated with ANN are summarized
in Table~3 for several ranges of spectral types, confirming that ANN
gives an accurate estimation of the Bayesian probabilities.  The bias
towards hot spectral types, intrinsic to the IUE archive, is also
evident in this sample.

\begin{table} \label{tab:prob} \caption[]{Probability distributions}
\begin{flushleft}
  \begin{tabular}{ccc}
    \hline\noalign{\smallskip}
    Class & Atlas & ANN \\
    \noalign{\smallskip}
    \hline\noalign{\smallskip}
    O0 - O4 & 0.048 & 0.056 \\
    O5 - O9 & 0.141 & 0.154 \\
    B0 - B4 & 0.322 & 0.291 \\
    B5 - B9 & 0.185 & 0.190 \\
    A0 - A4 & 0.132 & 0.141 \\
    A5 - A9 & 0.066 & 0.073 \\
    F0 - F4 & 0.075 & 0.063 \\
    F5 - F9 & 0.022 & 0.008 \\
    G0 - G4 & 0.004 & 0.007 \\
    \noalign{\smallskip}
    \hline
  \end{tabular} \end{flushleft} \end{table}

\section{Conclusions}

The relative performance of two automated classification methods have
been analyzed using IUE low-dispersion spectra of normal stars.  The
methods do not assume prior knowledge about the spectral types to be
classified and the algorithms can be applied directly to the observed
flux distributions. The analysis confirms the qualitative results
obtained on the same sample, by using multivariate statistics on
selected spectral features.

The simple metric distance gives a good classification, but the
results produced with the ANN classifier are better.  The accuracy
obtained with this method is 1.1 spectral subclasses.  ANN classifiers
have several advantages: they estimate Bayesian probabilities for each
spectral type and, in addition, it is possible to identify spectra
with uncertain classifications. The method is robust enough to be used
in case of spectra with missing information and further improvements
can be obtained with IUE spectra by rejecting bad pixels as indicated
in the quality flags associated with the spectral values.

This research will be continued to derive physical parameters by using
stellar models in connection with observed spectra.  Unsupervised ANN
classification will be used on the same set.

\begin{acknowledgements}
  We thank Dr. A. Cassatella and Dr. B. Montesinos for the useful
comments and suggestions.  In addition we are grateful to Dr. A. Heck
for the wise advise as referee of the article.  The work of E.F.V. has
been supported by the fellowship programme CAPES funded by the
Ministry of Science and Education of Brazil.  \end{acknowledgements}

\end{document}